\documentclass{article}

\def\ParSkip{} % uncomment to enable paragraph skip
\usepackage{amssymb,amsmath,amsthm,bbm}
\usepackage{verbatim,float,url,dsfont}
\usepackage{graphicx,subfigure,psfrag}
\usepackage{mathtools,enumitem}
\usepackage{multirow}
\usepackage{ragged2e}
\usepackage{xr-hyper}
\usepackage{array}

\usepackage[colorlinks=true,citecolor=blue,urlcolor=blue,linkcolor=blue]{hyperref}
\usepackage[margin=1in]{geometry}
\usepackage[round]{natbib}

\usepackage[utf8]{inputenc} % allow utf-8 input
\usepackage[T1]{fontenc}    % use 8-bit T1 fonts
\usepackage{booktabs}       % professional-quality tables
\usepackage{nicefrac}         % compact symbols for 1/2, etc.
\usepackage{microtype}      % microtypography
\usepackage{amsfonts}
\usepackage{authblk}
\usepackage{algorithm2e}
\usepackage{placeins}

\ifdefined\TimesFont 
\usepackage{times} % use times font
\fi

\ifdefined\ParSkip 
\usepackage{parskip} % use par skip
\fi

\newtheorem{theorem}{Theorem}
\newtheorem{lemma}{Lemma}

\newtheorem{proposition}{Proposition}
\theoremstyle{definition}
\newtheorem{remark}{Remark}

\newtheorem*{assumption*}{\assumptionnumber}
\providecommand{\assumptionnumber}{}


\makeatletter
\newcommand*\rel@kern[1]{\kern#1\dimexpr\macc@kerna}
\newcommand*\widebar[1]{%
  \begingroup
  \def\mathaccent##1##2{%
    \rel@kern{0.8}%
    \overline{\rel@kern{-0.8}\macc@nucleus\rel@kern{0.2}}%
    \rel@kern{-0.2}%
  }%
  \macc@depth\@ne
  \let\math@bgroup\@empty \let\math@egroup\macc@set@skewchar
  \mathsurround\z@ \frozen@everymath{\mathgroup\macc@group\relax}%
  \macc@set@skewchar\relax
  \let\mathaccentV\macc@nested@a
  \macc@nested@a\relax111{#1}%
  \endgroup
}
\makeatother

\title{Multi-label Classification under Uncertainty: A Tree-based Conformal Prediction Approach}
\author{Chhavi Tyagi}
\author{Wenge Guo\thanks{Author e-mail addresses: ct364@njit.edu, wenge.guo@njit.edu}}
\affil{Department of Mathematical Sciences\\
       New Jersey Institute of Technology,
       Newark, NJ 07102, USA}

\date{\today}

\begin{document}
\maketitle

\begin{abstract}
Multi-label classification is a common challenge in various machine learning applications, where a single data instance can be associated with multiple classes simultaneously. The current paper proposes a novel tree-based method for multi-label classification using conformal prediction and multiple hypothesis testing. The proposed method employs hierarchical clustering with labelsets to develop a hierarchical tree, which is then formulated as a multiple-testing problem with a hierarchical structure. The split-conformal prediction method is used to obtain marginal conformal $p$-values for each tested hypothesis, and two \textit{hierarchical testing procedures} are developed based on marginal conformal $p$-values, including a hierarchical Bonferroni procedure and its modification for controlling the family-wise error rate. The prediction sets are thus formed based on the testing outcomes of these two procedures. We establish a theoretical guarantee of valid coverage for the prediction sets through proven family-wise error rate control of those two procedures.  We demonstrate the effectiveness of our method in a simulation study and two real data analysis compared to other conformal methods for multi-label classification.

\end{abstract}

\section{Introduction}
\label{sec:intro}

Most machine learning research on classification deals  with $2n$ instances as $\{(X_j,Y_j )\}_{j=1}^{2n}$ with features $X_j \in \mathbb{R}^d $ and a response variable $Y_j \in \mathcal{Y} = \{1,\ldots,K\}. $ Each feature vector $X_j$ is associated with a single response $Y_j$. As opposed to this standard setting, in multi-label classification, each instance can belong to multiple classes, $Y_j = (Y_j^1,\ldots,Y^c_j)$ is the response vector for $j^{th}$ observation with $c$ labels. There are many real-world problems in which such a setting is natural. For instance, in medical diagnosis, a patient may be suffering from diabetes and prostate cancer simultaneously. Another example is the problem of semantic scene classification, where multiple class labels can describe a natural scene (e.g., mountain, sea, tree, sun etc.)
 
Many multi-label classification methods have been developed like Binary Relevance (BR) \citep{boutell2004learning}, classifier chains \citep{Read_2009} and Label Powerset (LP) \citep{tsoumakas2007multi,boutell2004learning} using black-box algorithms like logistic regression, svm etc. Given the growing utilization of black-box methods and the heightened risk of making erroneous decisions, it is crucial to develop techniques for accurately estimating the uncertain nature of their predictions.  For example, in the field of medical imaging, a multi-label classifier may be employed to recognize various conditions in a single scan, such as identifying the presence of a tumor and its level of malignancy. If the classifier makes an incorrect prediction, it could lead to misdiagnosis and subsequent incorrect treatment, which can have serious consequences for the patient.\\
\indent Therefore, it is essential for the classifier to provide not only an estimation of the most probable outcome but also a quantification of uncertainty that is actionable, such as a set of predictions that can be proven to contain the true diagnosis with a high level of confidence (for example, $90\%$). Conformal prediction (CP) is a popular method developed for uncertainty quantification. Importantly, conformal prediction is a distribution-free framework. This means that it does not rely on any assumptions about the distribution of the data or the model being used. Only a few works have been done to accommodate statistical guarantees using conformal prediction for multi-label classification like binary relevance multi-label conformal predictor (BR-MLCP) \citep{wang2015comparison} and label powerset multi-label conformal predictor (PS-MLCP1) \citep{papadopoulos2014cross} and \citep{wang2014reliable}. These procedures do not accommodate dependencies among the multiple response variables and are computationally inefficient for large number of response variables. Additionally, in multi-label classification, the response variables (i.e., the class labels) can often have missing information, i.e., with $c$ labels and $2^c$ possible labelsets, not all labelsets are present in the data. This can have significant implications for the accuracy and reliability of the predictions made by existing methods. In this paper, we develop a tree-based conformal prediction method for multi-label classification using multiple hypothesis testing. The method is not only able to have quantification of uncertainty with statistical guarantees but also takes into account of any missing information in the response variables. Since it uses a tree-based approach, the method is computationally efficient for large number of labels and considers label dependence. Thus, we have created a classification approach that delivers precise prediction sets with statistical coverage guarantee. \\
\indent The rest of the paper is organized as follows. With conformal prediction framework and related works discussed in section \ref{sec:cp}, we provide notations and describe our proposed tree-based procedure (\textit{TB}) in section \ref{sec:TB-MLCP}. The statistical properties of the procedure are analyzed in Section \ref{sec:statproperty}. In section \ref{sec:missing}, we discuss our proposed procedures developed to account for missing information. Numerical findings from the simulation study and real data analysis are presented in section \ref{sec:experiment}. We conclude the paper in section \ref{sec:discussion} with some remarks on the present work and brief discussions on some future research topics. We defer the proofs of the propositions in section \ref{sec:statproperty} to the appendix \ref{app:proofs}. 

\section{Background and related work}\label{sec:cp}

\subsection{Conformal Prediction}\label{cp}

Conformal prediction (CP) is a statistical learning framework that provides a prediction set $\hat{C}(X_{2n+1})$, instead of point predictions with statistical guarantees for a given test instance $X_{2n+1}$. 

Given dataset $ \mathcal{D} \equiv \{(X_i,Y_i)\}_{i=1}^{2n} \equiv \{Z_i\}_{i=1}^{2n} $ with $ X_i \in \mathcal{X}, \mathcal{X} \subset \mathbb{R}^{d} $  is input of real valued attributes and $Y_i \in \mathcal{Y} \in \{0,1\}$ , where $(X_{2n+1}, Y_{2n+1})$ and $\{(X_i, Y_i)\}_{i=1}^{2n}$ are drawn exchangeably from $P_{XY}$. The goal of CP is to construct a prediction set $ \hat{C}(X_{2n+1}) $ that contains $ Y_{2n+1} $ with probability at least $ (1 - \alpha),$ where $\alpha$ is user-defined significance level such that $\alpha \in (0,1).$ \\

\begin{figure}[h]
    \centering
    \includegraphics[width= \textwidth]{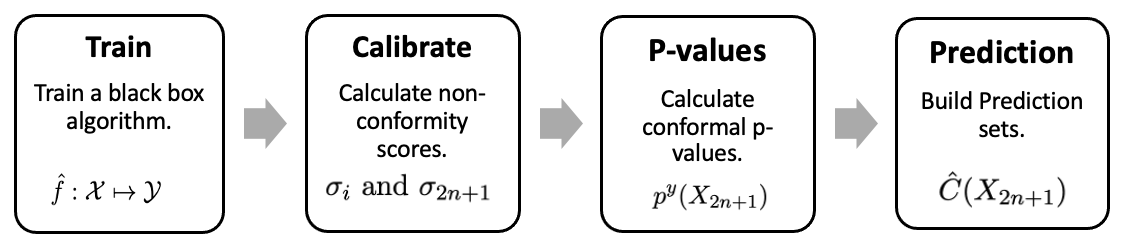}
    \caption{Steps of Conformal Prediction.}
    \label{fig:p1}
\end{figure}
\FloatBarrier

Figure \ref{fig:p1} summarizes the main steps involved in the split-conformal prediction framework, starting from training the model to the final prediction set. The initial step involves splitting the data into two parts, i.e., proper training $\mathcal{D}_{tr} \equiv \{(X_i,Y_i)\}_{i=1}^{n}$ and calibration $\mathcal{D}_{cal} \equiv \{(X_i,Y_i)\}_{i=n+1}^{2n}$. The first step involves training the model using a black-box algorithm mapping features $X$ to a real attribute $Y.$ The next step, calibration is a crucial component of the conformal prediction framework as it defines a function (called as \textit{non-conformity score}) that measures how well the true response value $y$ \textit{conforms} with the predictions of our fitted model $\hat{f}$. Given the score function $\sigma: \mathcal{X} \times \mathcal{Y} \rightarrow \mathbb{R}$, the non-conformity score for observations in the calibration set is given by:
 \begin{equation}
     \sigma_i = \sigma(X_i,Y_i), \hspace{2mm} i=n+1, \ldots, 2n
 \end{equation}
 
For a given test observation, $X_{2n+1},$ we compute the non-conformity score for each $y \in \mathcal{Y}$ as
\begin{equation}
     \sigma_{2n+1}^y = \sigma(X_{2n+1},y)
\end{equation}

In our specific case, we use the non-conformity score as 1-predicted probability i.e.,
\begin{equation}\label{eq:ncm_cal}
    \sigma_i = 1-\hat{f}(Y_i|X_i), i=n+1,\ldots,2n
\end{equation}
\begin{equation}\label{eq:ncm_test}
    \sigma_{2n+1}^y = 1-\hat{f}(y|X_{2n+1}), \text{ for each } y \in \mathcal{Y}
\end{equation}
\iffalse
\begin{enumerate}
\item  Set $Y_{n+1} =y$, and form augmented set $\mathcal{A} = \{Y_1,\ldots,Y_n,y_{n+1}\}$
   \item  Calculate non-conformity score, $\sigma_i = 1-\hat{f}(y,\mathcal{A}|X_{n+1})$, for $i=1,\ldots,n+1.$
\end{enumerate}
\fi

  The third step showcases to calculate conformal $p$-values as given in equation (\ref{eq:1}), which provides a way to quantify the confidence in the predictions
\begin{equation}\label{eq:1}
    p^{y}(X_{2n+1}) = \dfrac{ \sum_{i=n+1}^{2n} \{\mathbb{I}(\sigma_i < \sigma_{2n+1}^y) + U_{2n+1} \cdot \mathbb{I}(\sigma_i = \sigma_{2n+1}^y)\}}{n}
\end{equation}
where $U_{2n+1} \sim Unif(0,1)$ is random variable independent of $\sigma_i$'s introduced to break ties. The last panel illustrates to calculate prediction sets using the conformal $p$-values provided by equation (\ref{eq:2})
\begin{equation}\label{eq:2}
    \hat{C}(X_{2n+1}) = \{y \in \mathcal{Y}: p^{y}(X_{2n+1}) \geq \alpha\}
\end{equation}
 
 The prediction set constructed using the conformal $p$-values in equation (\ref{eq:1}) is subject to the guarantee of \textit{marginal coverage}, as expressed in Equation (\ref{eq:3}), which states that the probability of the true label of a new instance falling within the prediction set is at least $1 - \alpha$.
\begin{equation}\label{eq:3}
     \mathbb{P}(Y_{2n+1} \in \hat{C}(X_{2n+1})) \geq 1 - \alpha
\end{equation}

\iffalse
To address this limitation, we replace the split-conformal approach with cross-validation. Specifically, we use k-fold cross-validation to divide the data into k subsets, with each subset used once as the training, calibration, tuning and test set, respectively. 
\fi

\subsection{Related work}\label{sec:2.2}
Conformal prediction is a statistical framework that was first proposed by \cite{vovk1999machine, vovk2005algorithmic, shafer2008tutorial}. The literature on conformal prediction has identified two main types of conformal prediction methods, namely full conformal prediction (transductive) and split-conformal prediction (inductive) \cite{vovk2005algorithmic}, which is introduced in section \ref{cp}. Full conformal prediction involves using all the training observations to calculate the conformal p-values, which can become computationally expensive. On the other hand, split-conformal prediction is a hold-out method where part of the data is reserved for
training and the remaining are used for calibration, which may result in less accurate estimates of $\hat{f}$ than if all the data had been used for estimation but much computationally efficient. The framework has been widely used in regression problems \cite{romano2019conformalized, johansson2014regression, izbicki2019flexible, Gupta_2022, lei2014distribution} and classification problems \cite{10.5555/2671155, vovk2003mondrian, lei2014classification, sadinle2019least, romano2020classification}. A detailed introduction to the conformal prediction framework is provided in \cite{angelopoulos2022gentle}. Our work primarily builds on the use of split-conformal prediction \citep{vovk2005algorithmic}. Other types of conformal prediction methods include cross-conformal prediction \cite{vovk2015cross} and CV+/Jacknife+ \cite{barber2021predictive}.

The most relevant works in the context of multi-label classification problems using conformal prediction include Binary Relevance multi-label conformal predictor (BR-MLCP) \cite{wang2015comparison} and Power set multi-label conformal predictor (PS-MLCP1) \cite{papadopoulos2014cross} and \cite{wang2014reliable}. BR-MLCP uses a one-against-all method to decompose the multi-label dataset into multiple single-label binary classification problems. However, this method does not consider the dependency among the labels. PS-MLCP1 efficiently accounts for the dependency among the labels but does not consider any missing information on the labels and also runs into computational issues for large number of labels. \cite{cauchois2021knowing} developed a tree-structured method to consider the interaction among the labels. Another framework that addresses multi-label classification problems using conformal prediction is risk control introduced by  \cite{bates2021distribution, angelopoulos2022conformal}. \cite{bates2021distribution} proposed a method for constructing distribution-free prediction sets that control the risk of incorrect predictions by introducing a new loss function that directly controls the risk of incorrect predictions. Specifically, their approach minimizes the expected loss over the prediction sets, subject to the constraint that the coverage probability is at least as large as the specified level of significance. \cite{angelopoulos2022conformal} proposed a new method for controlling the risk of prediction sets, which is highly relevant for multi-label classification problems in the field of conformal prediction. Their approach, called Conformal Risk Control (CRC), uses a novel loss function to optimize the trade-off between coverage and size of the prediction sets.

In contrast to traditional conformal prediction methods, CRC aims to minimize the overall risk of incorrect predictions by taking into account the cost of making errors.

There are also several statistical extensions to conformal prediction, including the ideas of non-exchangeability like time series by \cite{xu2021conformal}, outlier testing by \cite{bates2022testing}, covariate shift by \cite{tibshirani2019conformal}, and image classifiers by \cite{angelopoulos2021uncertainty}. 

\section{Proposed Tree-based multi-label Conformal Prediction Method (\textit{TB})}\label{sec:TB-MLCP}

\subsection{Notations}
    Consider the data set $ \mathcal{D} = \{(X_j,{Y}_j)\}_{j=1}^{2n} =\{Z_j\}_{j=1}^{2n}, $ with $ X_j \in \mathcal{X}, \mathcal{X} \subset \mathbb{R}^{d} $  is input of real valued attributes and  $ {Y}_j = (Y^{1}_j,\ldots,Y^{c}_j) $ is the response vector for $j^{th}$ observation. Each $Y_j^i \in \{0,1\}$ denotes the label of the $i^{th}$ class in the $j^{th}$ observation, which is labeled as $1$ if the instance belongs to the $i^{th}$ class and $0$ otherwise.
     Let $ (X_{2n+1},{Y}_{2n+1}) $ be the test observation. The goal is to provide prediction set $\hat{C}(X_{2n+1})$ for unobserved ${Y}_{2n+1}$ based on a given test instance $X_{2n+1}$ to show the  statistical guarantee on the prediction, given a significance level $\alpha \in (0,1)$. 
    
\subsection{Procedure}
Our proposed procedure consists of five main steps. Firstly, a hierarchical tree is developed based on all possible labelsets obtained from $c$ labels. Secondly, we formulate a hypothesis for each node of the tree. Since we have multiple nodes on the tree, it is a multiple hypothesis testing (MHT) problem with hierarchical structure. Thirdly, we compute conformal $p$-values for each tested hypothesis using the split-conformal prediction method and any black-box algorithm. Fourthly, we develop two types of \textit{hierarchical testing procedures} based on conformal $p$-values for controlling family-wise error rate (FWER). Finally, we form the prediction set from the test outcomes of the previous step by leveraging split-conformal prediction.  
     
\subsubsection{Hierarchical Tree}\label{sec:tree}
       
    In this section, we present a method for constructing a hierarchical tree for multi-label classification by utilizing hierarchical clustering on the binary representation of all possible labelsets, denoted by ${s_i} \in \{0,1\}^c, i=1,\ldots,2^c$. The use of hierarchical clustering is motivated by the need to handle a large number of labels in multi-label classification and to reduce the number of labelsets used for classification while preserving an elegant and manageable tree structure.

    To build the hierarchical tree, we need to define a distance metric and linkage criteria between the labelsets. For our purposes, we use the Hamming distance \citep{6421371}, which is a widely used metric in the binary setting, to calculate the dissimilarity, $d_{ij}$, between each pair of labelsets ${s_i}$ and ${s_j}$, for $i, j = 1, 2, ..., l$. Given two binary vectors ${s_1}$ and ${s_2}$ with $c$ elements each, the Hamming distance between them is defined as:

\begin{equation}\label{eq:hamming_dist}
    d_{H}({s_1}, {s_2}) = \sum\limits_{i=1}^c \mathbb{I}\{s_{1i} \neq s_{2i}\}
\end{equation} 
where $s_j = (s_{j1},\ldots,s_{jc})$, for $j=1,2.$ Thus, the Hamming distance counts the number of mismatches between two binary vectors.

    To calculate the distance among the clusters of labelsets, we use the complete linkage method \citep{6421371}. In complete linkage, the distance between any two clusters $C_K$ and $C_L$ is defined as the maximum distance between two observations that belong to different clusters,
\begin{equation}\label{eq:complete_link}
    D_{KL} = \max\limits_{{s \in C_{K}, t \in C_{L}}} d(s,t).
\end{equation}
    
    The advantage of using complete linkage is that it is less sensitive to noise and outliers compared to other linkage methods such as single linkage or centroid linkage. The steps for constructing a hierarchical tree are outlined in Algorithm \ref{alg:alg1}.

\RestyleAlgo{ruled}
\SetKwComment{Comment}{/* }{ */}

\begin{algorithm}
\caption{Hierarchical Clustering}
\label{alg:alg1}
 % older versions of algorithm2e have \dontprintsemicolon instead
 % of the following:
 %\DontPrintSemicolon
 % older versions of algorithm2e have \linesnumbered instead of the
 % following:
 %\LinesNumbered
\KwIn{Labelsets ${s_i} \in \{0,1\}^c,$ $i=1,\ldots,2^c$}
\KwOut{Hierarchical Tree}
{\textit{Cluster size} $\leftarrow 2^c$. \tcp*[f]{Each labelset is a cluster.}\\
Initialize a hierarchical tree $T$ with $2^c$ leaves, one for each labelset.\\
Compute pairwise distances $(d_{ij})$ between labelsets ${s_i},{s_j}$ as defined in equation (\ref{eq:hamming_dist}) and store in a matrix.}\\
\While{Cluster size $ > 1$}
{Find the pair of nodes with the smallest distance from the matrix formed.\\
    Merge the two nodes into a new parent node, and add edges to connect the parent node to the two children.\\
     \tcp*[f]{Reduce number of clusters by 1.}\\
    Compute the distance between the parent node and all other nodes in $T$ from equation (\ref{eq:complete_link}).\\
}
\end{algorithm}

\begin{figure}[h]
    \centering
    \includegraphics[scale=0.5]{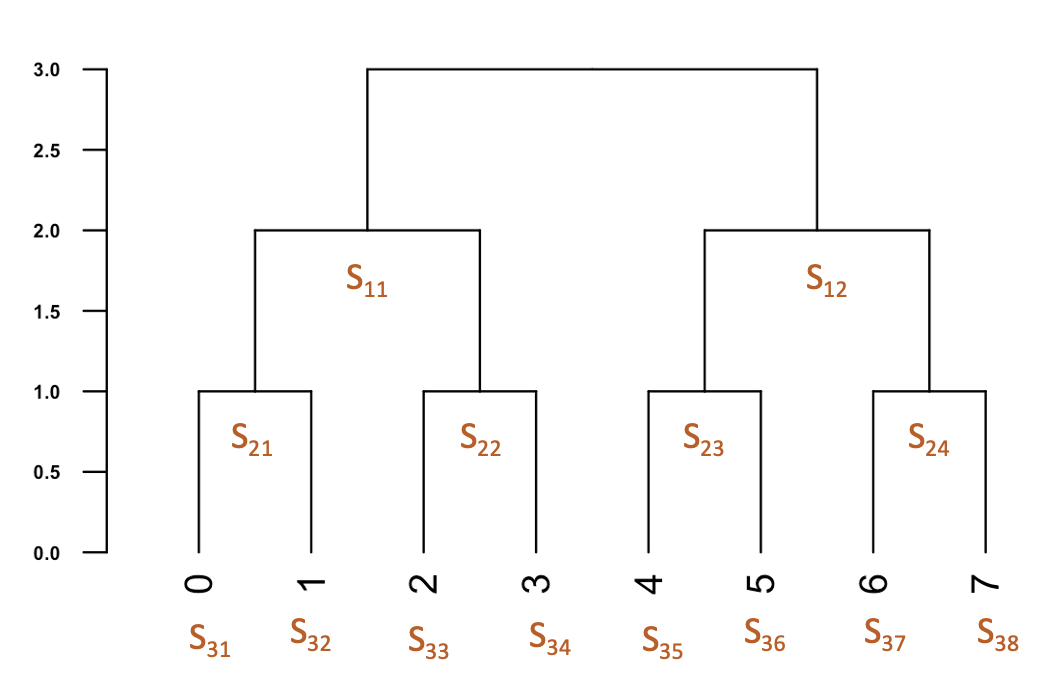}
    \caption{Hierarchical tree with $c=3$ labels.}
    \label{fig:3label_tree}
\end{figure}

In Figure \ref{fig:3label_tree}, we illustrate the hierarchical tree for the case of 3 labels. In this tree, each node represents a subset of the labelset space. We use the terms ``parent'' and ``children''  to describe the relationships between nodes in the tree. The parent of a node is the immediate ancestor in the tree, and its children are the immediate descendants. For example, if we select node $S_{21}$ from the tree, its parent would be $S_{11}$ which contains the labelset $\{0,1,2,3\}$, and its children would be $S_{31}$ and $S_{32}$ which contain the labelsets $\{0,1\}$ and $\{2,3\}$ respectively. This terminology describes the labelsets contained within the parent and children nodes in decimal form (i.e., base 10).  Specifically, the parent node contains a larger subset of labelset than its children nodes.  The depth of a hierarchical tree is defined as the number of layers in the tree, where the root node is at layer 0 and the leaf nodes are at the maximum layer. Each layer consists of nodes with the same depth. As an illustration, in Figure \ref{fig:3label_tree}, node $S_{11}$ and $S_{12}$ are located at depth 1, and all 8 leaf nodes are at the maximum depth of 3. Nodes at depth 0 do not have parents and nodes at the maximum depth of the tree (i.e., leaf nodes) do not have children.

\FloatBarrier

\begin{remark}
    In the hierarchical tree, the union of all nodes in a layer covers the entire response space, ${Y}$, i.e., for each layer $i=1,\ldots,c$, we have $\cup_{k=1}^{n_i} S_{ik} = {Y}$, where $n_i$ is the number of nodes in $i^{th}$ layer. Additionally, for each pair of distinct nodes  $S_{ik}$ and  $S_{ik'} $ in the same layer $i$, their intersection is empty, i.e., $S_{ik} \cap S^{'}_{ik} = \phi$.
\end{remark}

\subsubsection{Multiple Hypothesis Testing Problem}\label{sec:mht}
 
In this section, we formulate the multi-label classification problem as a multiple hypothesis testing (MHT) problem based on the hierarchical tree established in Section \ref{sec:tree}. For each node in the tree, we construct a hypothesis. Since we have multiple nodes in the tree, it becomes a multiple hypothesis testing (MHT) problem. The hypotheses are constructed such that under the null hypothesis, the test label belongs to the subset of labelsets corresponding to that node; under the alternative hypothesis, the test label does not belong to the subset of labelsets corresponding to that node. 

We construct the hypothesis layer by layer, starting from the second layer (i.e., depth 1) and proceeding until the leaves of the tree (i.e., maximum depth). 
For each layer $i$, $i=1,\ldots,c$, assume $n_i$ as the number of nodes in each layer. The structure for constructing these hypotheses is discussed below.

\noindent \textit{$i^{th}$ Layer:} Layer $i$ corresponds to the depth $i$ on the tree in Figure \ref{fig:3label_tree}. Then the hypotheses are as follows:
\begin{center}\label{eq:}

$H_{i1}: {Y_{2n+1}} \in \mathcal{S}_{i1}  \text{ versus }  H^{'}_{i1}: {Y_{2n+1}}  \notin \mathcal{S}_{i1} $

\

$ H_{i2}: {Y_{2n+1}} \in \mathcal{S}_{i2} $ versus $ H^{'}_{i2}: {Y_{2n+1}} \notin \mathcal{S}_{i2} $

\hspace{2mm}{\vdots}

$ H_{in_i}: {Y_{2n+1}} \in \mathcal{S}_{in_i} $ versus $ H^{'}_{in_i}: {Y_{2n+1}} \notin \mathcal{S}_{in_i} $

\end{center}

\noindent This procedure is repeated until we reach the tree's leaf layer, i.e., $i=c.$

\noindent \textit{Mathematical properties of the constructed hypotheses:}

\begin{enumerate}
    \item Due to the structural relationship among the nodes of the tree, in each layer $i$, exactly one hypothesis is true.
    \item If a children hypothesis is true, then its parent hypothesis is also true. Conversely, if a parent hypothesis is false, then its children's hypothesis are automatically false. 
    \end{enumerate}

\subsubsection{Conformal p-values}\label{sec:conformal_pvalues}
This section describes the method to compute the conformal $p$-values for the hypotheses constructed in Section \ref{sec:mht}. First, the data is split into training, calibration and tuning data as $\mathcal{D}_{tr}$, $\mathcal{D}_{cal}$ and  $\mathcal{D}_{tune}$ respectively using split-conformal prediction. Then, $\mathcal{D}_{tr}$ and $\mathcal{D}_{cal}$ are transformed from $(X_j,Y_j) $ to $(\tilde{X}_j,\tilde{Y}_j)$ for each layer $i$ as described below.
\begin{equation}\label{eq:4}
   \tilde{X}_j=X_j, \text{ and } \tilde{Y}_j= k, \text{ if } Y_j \in S_{ik},  
\end{equation}
where $k=1,2,\ldots,n_i.$ Thus, each layer is transformed as a multi-class classification problem with transformed data as 
${\tilde{\mathcal{D}}}_{tr}$ and ${\tilde{\mathcal{D}}}_{cal}$. Algorithm \ref{alg:pvalues} describes the general procedure to calculate conformal $p$-values for each node on the tree using split-conformal prediction, which includes tuning set, $\mathcal{D}_{tune}$ used later in Algorithm \ref{alg:algo} for tuning parameter $\lambda^*.$

In addition to the procedure described above, there are alternative approaches to computing conformal $p$-values. One such approach involves transforming the problem into multiple multi-class classification problems for each layer. For instance, if nodes in a given layer share the same parent, a multi-class classification transformation is applied to those nodes. This alternative approach provides an alternate way to compute conformal $p$-values and may interest researchers seeking to explore different methods for solving this problem.

\subsubsection{Controlling FWER}\label{sec:FWER}
In multiple-testing problems, control of some overall error rate such as the family-wise error rate (FWER) and the false discovery rate (FDR) \citep{hochberg1987multiple}, is crucial. 

We use a \textit{hierarchical testing procedure} to test the hypotheses formulated in section \ref{sec:mht}. Tree-based constructed hypotheses $H_{ik}$ are tested layer by layer with critical value $\alpha_i$ for $i^{th}$ layer and $k^{th}$ node, $i=1,\ldots,c$ and $k=1,\ldots,n_i$ such that $\sum_{i=1}^c\alpha_i \leq \alpha.$ For each layer $i$, if $p^{i,k} < \alpha_i$, we reject the hypothesis and also reject all its children hypotheses and move to the next layer until the leaf layer. 

Our proposed procedure focuses on controlling FWER, which is defined as the probability of not making any Type I error. It implies we do not allow making any Type I error. We consider the following two specific \textit{hierarchical testing procedures} for controlling the family-wise error rate (FWER).

\begin{enumerate}
    \item \textbf{Procedure 1}: In this \textit{hierarchical testing procedure}, the significance level for each layer is set to $\alpha_i =  \alpha/c$. This method is a Bonferroni procedure for hierarchical testing. It is known to be relatively less powerful.
    
    \item \textbf{Procedure 2}: In this \textit{hierarchical testing procedure}, the significance level for each layer is set to $\alpha_i=\lambda \alpha$, where $\lambda$ is the tuning parameter. We use Algorithm \ref{alg:algo} to perform hyperparameter tuning using binary search and find the optimal value of $\lambda$, denoted as $\lambda^*$ s.t. $\alpha_i= \lambda^* \alpha$, which helps to overcome the conservativeness of the first method.
\end{enumerate} 

\noindent Algorithm \ref{alg:testing} outlines the steps involved in implementing the FWER control procedure.

\begin{remark}
     In Algorithm \ref{alg:testing}, to utilize the first procedure of FWER control, we replace $\tilde{\alpha_i}$ with $\alpha_i = \alpha/c$ and there is no need for $\mathcal{D}_{tune}$. If the second procedure is being employed, we substitute $\tilde{\alpha_i}$ with $\alpha_i = \lambda^* \alpha$.
\end{remark}

\RestyleAlgo{ruled}
\SetKwComment{Comment}{/* }{ */}

\begin{algorithm}[h]
\caption{Conformal $p$-values}\label{alg:pvalues}
\KwIn{$\mathcal{D} = \{(X_j,{Y}_j)\}_{j=1}^{2n}$, test observation $X_{2n+1},$ black-box algorithm $\mathcal{B}.$}
\KwOut{Conformal p-values: $p^{i,k}_{tune}(X_l)$ and $p^{i,y}_{test}(X_{2n+1})$.}
Randomly split the data into 3 equal disjoint subsets, $\mathcal{D}_{tr}$, $\mathcal{D}_{cal}$ and $\mathcal{D}_{tune}$ with $n_1,n_2,n_3$ observations in each set respectively s.t. $n_1+n_2+n_3=2n$. \\
\ForAll {$i \in \{1, 2, \ldots, c $\}}{
    Obtain ${\tilde{\mathcal{D}}}_{tr}$ and ${\tilde{\mathcal{D}}}_{cal}$ for each layer as in equation (\ref{eq:4}).\\
    Train $\mathcal{B}$ on all samples in ${\tilde{\mathcal{D}}}_{tr}$, $\hat{f}: \mathcal{B}(\{(X_j,Y_j)\}_{j \in {\tilde{\mathcal{D}}}_{tr}} )$\\
\ForAll {$k \in \{1, 2, \ldots, n_i $\}}{    
    Calculate the non-conformity scores for $i^{th}$ layer, $k^{th}$ node, and $j^{th}$ observation on $\tilde{\mathcal{D}}_{cal}$ as $\sigma^{ik}_{cal}(X_j), j=1,\ldots,n_2$ using equation (\ref{eq:ncm_cal}).\\
    Calculate the non-conformity scores for $i^{th}$ layer, $k^{th}$ node, and $l^{th}$ observation on $\tilde{\mathcal{D}}_{tune} $ as $\sigma^{ik}_{tune}(X_l),$ $l=1,\ldots,n_3$ using equation (\ref{eq:ncm_cal}).\\
    Given $(X_l,Y_l) \in \mathcal{D}_{tune},$ conformal $p$-values for $i^{th}$ layer, $k^{th}$ node and $l^{th}$ observation is,\\
    \begin{equation}\label{eq:5}
        p^{i,k}_{tune}(X_l) = \dfrac{ \sum_{j=1}^{n_2} \{\mathbb{I}(\sigma_{cal}^{ik}(X_j) < \sigma_{tune}^{ik}(X_l)) + U_{n_2+1} \cdot \mathbb{I}(\sigma_{cal}^{ik}(X_j) = \sigma_{tune}^{ik}(X_l))\}}{n_2+1},
    \end{equation}  
    Given $X_{2n+1} \in \mathcal{D}_{test}$ and $y=k$, we form $(X_{2n+1},y)$, compute the non-conformity score $\sigma^{i,y}_{test}(X_{2n+1})$ from equation (\ref{eq:ncm_test}) and corresponding conformal $p$-value as\\
    \begin{equation}\label{eq:6}
        p^{i,y}_{test}(X_{2n+1}) = \dfrac{\sum_{j=1}^{n_2}\{\mathbb{I}(\sigma_{cal}^{ik}(X_j) < \sigma_{test}^{i,y}(X_{2n+1})) + U_{n_2+1} \cdot \mathbb{I}(\sigma_{cal}^{ik}(X_j) = \sigma_{test}^{i,y}(X_{2n+1}))\}}{n_2+1}
    \end{equation}
}
}
\end{algorithm}

\RestyleAlgo{ruled}
\SetKwComment{Comment}{/* }{ */}

\begin{algorithm}[h]\label{alg:testing}
\KwIn{$\alpha \in (0,1), \tilde{\alpha_i}$, $i=1,2,\ldots,c$ s.t. $\sum_{i=1}^c  \tilde{\alpha_{i}} \leq \alpha.$}
\KwOut{$\mathcal{S}(\alpha)$: Set of accepted hypotheses in the leaf layer.}
\While{\text{layer, i} $< c$}{
\For{k $\in \{1,\ldots,n_i\}$}{
    %$\alpha_i = w_i \alpha/c$\; 
    For $y=k$, compute conformal $p$-values for each node $y$ in the $i^{th}$ layer as $p^{i,y}(X_{2n+1})$, using Algorithm \ref{alg:alg1}.\\ 
    Test $H_{ik},$ in layer $i$ at $\tilde{\alpha_i}$.\\
    Reject $H_{ik}$ and all its children's hypotheses if $p^{i,y}(X_{2n+1}) < \tilde{\alpha_i}$.\\ 
}
}
For layer $i=c$ and $k=1,\ldots,n_i$, Reject $H_{ik}$ if $p^{i,y}(X_{2n+1}) < \tilde{\alpha_i}$.\\
\caption{FWER controlling procedure for MHT}
\end{algorithm}
\FloatBarrier

{\small
\begin{algorithm}[h]
\caption{Hyper-parameter tuning $\lambda^*$}\label{alg:algo}
\KwIn{ $\alpha_{f} = 0$,  $\alpha_{c}= \alpha$, $\alpha^* =( \alpha_{f} +  \alpha_{c} )/2$, $\alpha \in (0,1).$}
\KwOut{$\lambda^*$}
\While {$( \alpha^* > 0 \text{ and } \alpha^* \leq \alpha )$}{
Compute prediction set for each observation $(X_j,Y_j)\in \mathcal{D}_{tune}, j=1,\ldots,n_3$  is given by
\begin{equation}
    \hat{C}_{tune}(X_j,\alpha^*) = \{ y : \text{Labelset y for which leaf hypothesis } \in \mathcal{S}(\alpha^*) \}
\end{equation} \tcp*[f]{This is done using output of Algorithm \ref{alg:testing} where $X_{2n+1}$ is replaced by $X_j \in \mathcal{D}_{tune}.$}

Calculate the coverage on $ \hat{C}_{tune}(X_j,\alpha^*)$ as 
$$\mathcal{C}(\alpha^*) = \dfrac{1}{n_3}\sum_{j=1}^{n_3}\mathbb{I}\left\{Y_j \in \hat{C}_{tune}(X_j,\alpha^*)\right\} $$

\If{$\mathcal{C}(\alpha^*) < 1 - (1+1/n)(\alpha-1/n)$}{ $  \alpha_{c} \gets \alpha^*;$
 $\alpha^* =( \alpha_{f} +  \alpha_{c} )/2$}
 \If{$\mathcal{C}(\alpha^*) > 1 - (1+1/n)(\alpha-1/n)$}
    { $\alpha_{f} \gets \alpha^*;$
    $\alpha^* =( \alpha_{f} +  \alpha_{c} )/2$}
 \If{ $0 \leq \mathcal{C}(\alpha^*) \leq 1/n$ } {\textbf{break}  } 
}
$\lambda^*= \alpha^*/\alpha$
\end{algorithm}
}

\subsubsection{Prediction Set}\label{sec:pre_set}
To construct the prediction set, we utilize the output of Algorithm \ref{alg:testing}, which provides $\mathcal{S}(\alpha)$ as the set of accepted hypotheses in the leaf layer. The prediction set includes all labels for which the corresponding leaf hypotheses are not rejected at level $\alpha$.
The prediction set for $\mathcal{D}_{test} = X_{2n+1}$ is given by:
\begin{equation}\label{eq:pre_set}  
\hat{C}_{test}(X_{2n+1},{\alpha}) = \{ y : \text{Labelset } y \text{ for which leaf hypothesis } \in \mathcal{S}(\alpha)\},
\end{equation}
\iffalse
\begin{remark}
    In equation (\ref{eq:pre_set}), the value of ${\alpha}$ depends on the FWER control method used. If the first method described in section \ref{sec:FWER} is used, then ${\alpha} =  w_c\alpha/c$, where $w_c$ is the weight assigned to the leaf layer. On the other hand, if the second method of FWER control is deployed, then ${\alpha}=w_c\lambda^*\alpha$, where $\lambda^*$ is the optimal tuning parameter determined using Algorithm \ref{alg:algo}. 
\end{remark}

We present two implementations of our proposed procedure. The first implementation uses the split-conformal method, while the second implementation uses cross-validation. Both implementations follow the same five-step procedure described in Section \ref{sec:tree}-\ref{sec:pre_set}. The difference between the two methods lies in how the training and test sets are formed.
\fi
\

\paragraph{Split-Conformal Prediction Set:} Algorithm \ref{alg:split_final} outlines our proposed method, Tree-Based Multi-Label Conformal Prediction (\textit{TB}) with split-conformal method. For simplicity, we divide the data into three equal subsets, namely $\mathcal{D}_{tr}, \mathcal{D}_{cal} \text{ and } \mathcal{D}_{tune}.$ However, in practice, we can change the splitting ratio for better results. 

\

{\small

 \begin{algorithm}[H]
\caption{Split-Conformal $\lambda^*$ TB}\label{alg:split_final}
\KwIn{ $\mathcal{D} = {(X_j,{Y}_j)}_{j=1}^{2n}$, $\mathcal{D}_{test} = X_{2n+1}$, black-box $\mathcal{B}$, number of labels $c$.}
\KwOut{Prediction set, $\hat{C}_{test}(X_{2n+1},\tilde{\alpha})$ for the unobserved label $Y_{2n+1}$.} %and coverage, $\mathcal{C}(\tilde{\alpha})$.}
Build a hierarchical tree using Algorithm \ref{alg:alg1}.\\
{Randomly split the data into 3 equal subsets, $\mathcal{D}_{tr},\mathcal{D}_{cal}$ and $\mathcal{D}_{tune}.$}\\
Calculate the conformal $p$-values $p^{i,k}_{tune}(X_j)$ on the nodes of the tree for the labeled data in $\mathcal{D}_{tune}$ using Algorithm \ref{alg:pvalues}.\\
Perform multiple hypothesis testing (MHT) on the conformal $p$-values, $p^{i,k}_{tune}(X_j)$ using Algorithm \ref{alg:testing}.\\
Find the tuning parameter $\lambda^*$ using $p^{i,k}_{tune}(X_j)$ in Algorithm \ref{alg:algo}.\\
Calculate the conformal $p$-values $p^{i,y}_{test}(X_{2n+1})$ on the test data in $\mathcal{D}_{test}$ using Algorithm \ref{alg:pvalues}.\\
Perform multiple hypothesis testing (MHT) on the conformal $p$-values $p^i_{test}(X_{2n+1})$ with $\lambda^*$ using Algorithm \ref{alg:testing}.\\
Build the prediction set, $\hat{C}_{test}(X_{2n+1},\tilde{\alpha})$ for $\mathcal{D}_{test} = X_{2n+1}$ using equation (\ref{eq:pre_set}).\\
\iffalse
Calculate the coverage on $ \hat{C}_{test}(X_{2n+1},\tilde{\alpha})$ as 
$$\mathcal{C}(\tilde{\alpha}) = \mathbb{I}\left\{Y_{test}(X_{2n+1}) \in \hat{C}_{test}(X_{2n+1},\tilde{\alpha})\right\} $$
\fi
\end{algorithm}
}
\

\begin{remark}
    In Algorithm \ref{alg:split_final}, if we use procedure 1 of FWER control, then $\mathcal{D}_{tune}$ and calculation of $\lambda^*$ is omitted.
\end{remark}

\section{Statistical Property}\label{sec:statproperty}
 For any non-conformity score function, given the non-conformity scores of calibration data $(X_i,Y_i)_{i=1}^n$ as $\sigma_i$, $i=1,\ldots, n$ and for test data $(X_{n+1},Y_{n+1})$, where $Y_{n+1}$ is unknown as $\sigma_{n+1}$, we define smoothed conformal $p$-value for test observation as $p(X_{n+1},Y_{n+1})$ defined below:

 \begin{equation}\label{eq:pvalue}
    p(X_{n+1},Y_{n+1}) = \dfrac{ \sum_{i=1}^{n} \{\mathbb{I}(\sigma_i < \sigma_{n+1}^{Y_{n+1}}) + U_{n+1} \cdot \mathbb{I}(\sigma_i = \sigma_{n+1}^{Y_{n+1}})\}}{n+1}, 
 \end{equation}
 where $U_{n+1} \sim Unif(0,1)$ is random variable independent of $\sigma_i$'s introduced to break ties.
For the sake of simplicity, we will use the notation $\sigma_{n+1}$ to represent $\sigma_{n+1}^{Y_{n+1}}$ throughout the remaining proofs in Section \ref{sec:statproperty} and appendix \ref{app:proofs}.

\addtocounter{theorem}{-3}

\begin{proposition}[Validity of conformal $p$-values]\label{prop:valid}
 Given $(n+1)$ observations, $(X_i,Y_i)_{i=1}^{n+1}$ are exchangeable, then the smoothed conformal $p$-value defined in equation (\ref{eq:pvalue}) is uniformly distributed on the interval $(0, 1)$ i.e., for any given $t \in [0,1],$ we have
 $$\mathbb{P}\{p(X_{n+1},Y_{n+1}) \leq t\} = t$$

 \iffalse
 \begin{enumerate}
     \item The conformal $p$-values defined in equation (\ref{eq:5}) and (\ref{eq:6}) are valid i.e. satisfies
$$ \mathbb{P}\{p(\sigma_{n+1}) \leq t\} \leq t ,$$
 for any $t \in [0,1].$ 
 \item Moreover, if the non-conformity scores $\sigma_i$ are almost surely distinct, then the conformal $p$-values also satisfies
 $$\mathbb{P}\{p(\sigma_{n+1}) \leq t\} \geq t - \dfrac{1}{n+1}$$
 for any $t \in [0,1].$ \\
 \item Specifically, the conformal $p$-values are uniformly distributed between $\left\{\dfrac{1}{n+1},\ldots,1\right\}.$
\end{enumerate}
\fi

\end{proposition}

\begin{proposition} \label{prop:fwer_pr1} Under the same exchangeability assumption as in Proposition \ref{prop:valid}, any \textit{hierarchical testing procedure} defined in section \ref{sec:FWER} with critical values $\alpha_i$ satisfying $\sum_{i=1}^c \alpha_i  \leq \alpha$ strongly controls FWER at level $\alpha.$
 
\end{proposition}
\begin{remark}
      By using the result of Proposition \ref{prop:fwer_pr1}, procedure 1 described in section \ref{sec:FWER} with critical value $\alpha_i=\alpha/c$ strongly controls FWER at level $\alpha.$
  \end{remark}

To show the FWER control of procedure 2, we need to use the following lemma. 

\begin{lemma}
\label{lemma:label1}
\citep{romano2019conformalized} Suppose $Z_1,\ldots,Z_n$ are exchangeable random variables. For any $\alpha \in (0,1),$
$$\mathbb{P}\{Z_{n+1} \leq \hat{Q}_n((1+\dfrac{1}{n})\alpha)\} \geq  \alpha $$
Moreover, if the random variable $Z_1,\ldots,Z_n$ are almost surely distinct, then also,
$$\mathbb{P}\{Z_{n+1} \leq \hat{Q_n}((1+\dfrac{1}{n})\alpha)\} \leq  \alpha + \dfrac{1}{n} $$
\end{lemma}

\begin{proposition}\label{prop:fwer_pr2}  Under the same exchangeability assumption as in Proposition \ref{prop:valid}, procedure 2 defined in Section \ref{sec:FWER}, ensures strong control of FWER at level $\alpha$ i.e,
$$ FWER \leq \alpha$$
Moreover, the FWER is bounded from below by $\alpha -1/n$, i.e.,
$$ FWER \geq \alpha - \dfrac{1}{n}$$

\end{proposition}

\begin{remark}
    Proposition \ref{prop:fwer_pr2} shows procedure 2  is almost optimal in the sense that the corresponding FWER is almost equal to the pre-specified level $\alpha$ if $n$ is large enough.
\end{remark}

\begin{theorem}[Validity of Prediction Set]\label{theorem:pre_Setvalid} Under the same exchangeability assumption as in Proposition \ref{prop:valid}, the prediction sets obtained based on procedure 1 and 2 and equation (\ref{eq:pre_set}) are both guaranteed to be marginally valid, i.e., 
$$\mathbb{P}\{Y_{n+1} \in \hat{C}_{test}(X_{n+1},{\alpha})\} \geq 1 - \alpha.$$
Moreover, the prediction set based on procedure 2 also satisfies
$$\mathbb{P}\{Y_{n+1} \in \hat{C}_{test}(X_{n+1},{\alpha})\} \leq 1 - \alpha + \dfrac{1}{n}.$$

\begin{proof}
By using the same argument as in Proposition \ref{prop:fwer_pr2}, the event of $Y_{n+1} \in \hat{C}_{test} (X_{n+1},\alpha^*)$ is equivalent to  the only true null hypothesis is not rejected for each layer $i$, which in turn implies
\begin{equation}\label{eq:pre_val}  
 \mathbb{P}\{Y_{n+1} \in \hat{C}_{test}(X_{n+1},{\alpha})\} = 1- FWER\\       
\end{equation}
By using proposition \ref{prop:fwer_pr1} and \ref{prop:fwer_pr2}, we have marginal validity of prediction sets from both the procedures,
\begin{equation}\label{eq:v1}
     \mathbb{P}\{Y_{n+1} \in \hat{C}_{test}(X_{n+1},{\alpha})\} \geq 1 - \alpha.
\end{equation}
From equation (\ref{eq:pr2l_b}) of proposition \ref{prop:fwer_pr2}, we get the upper bound of the prediction set for Procedure 2 as
\begin{equation}\label{eq:v2}
     \mathbb{P}\{Y_{n+1} \in \hat{C}_{test}(X_{n+1},{\alpha})\} \leq  1-\alpha+\dfrac{1}{n}.
\end{equation}
This completes the proof.
\end{proof}
\end{theorem}

\section{Missing Information}\label{sec:missing}
In practical applications, it is common to encounter situations where not all possible labelsets are present in the data. This missing information can pose a challenge when implementing our proposed Tree-Based Multi-label Conformal Prediction (\textit{TB}) procedure described in section \ref{sec:TB-MLCP}. To address this issue, we have developed two modified versions of our procedure that can handle missing labelsets in different ways.
\begin{enumerate}
    \item The first approach referred to as \textit{Tree-Based MLCP with present labelsets (\textit{TB1})}, involves building the hierarchical tree using only the labelsets that are present in the data. Next, we formulate the multiple hypothesis testing (MHT) problem based on the resulting hierarchical tree. Finally, we apply Algorithm \ref{alg:split_final} to build a prediction set for an unobserved test instance. This approach assumes that the missing labelsets are simply absent from the data and does not attempt to make any assumptions about them. \cite{papadopoulos2014cross} and \cite{wang2014reliable} proposed Power-Set MLCP1 (PS-MLCP1), which uses a similar approach for handling missing labelsets.
    \item The second approach referred to as \textit{Tree-Based MLCP with parent p-value labelsets (\textit{TB2})}, addresses the issue of missing labelsets by using the hierarchical tree of all possible labelsets and assigning parent $p$-value to the nodes with missing information. Finally, we apply Algorithm \ref{alg:split_final} to build a prediction set for an unobserved test instance. This approach provides a more structured way of making predictions on missing labelsets and takes into account the relationships between different labelsets in the hierarchical tree. Due to the logical relationship between parent and children from the hierarchical tree, we can assign the parent $p$-value to the nodes with missing labelsets (i.e., missing information). Hence, it can handle the missing information more efficiently than PS-MLCP2 (refer section \ref{sec:experiment} for more details), where we use PS-MLCP1 to the present labelsets and add all missing labelsets in the predictions set to account for missing information.
\end{enumerate}

Overall, these two modified versions of our \textit{TB} procedure provide more robust ways of handling missing labelsets in real-world applications. The choice of approach will depend on the specific context and nature of the missing labelsets, and both approaches have their respective strengths and limitations. \textit{TB2} accounts for the missing information by assigning parent $p$-value to the nodes with missing information. However, this method is computationally intensive. On the other hand, \textit{TB1} will be computationally efficient, but it does not account for missing information in the data. A trade-off between the length of prediction sets and coverage rate arises when choosing between \textit{TB1} and \textit{TB2} methods. \textit{TB1}, which employs present labelsets, yields shorter prediction sets with low coverage compared to \textit{TB2}, which provides larger prediction sets with higher coverage.

\section{Experiments}\label{sec:experiment}
 In this section, we evaluate the performance of our proposed method, \textit{TB}, described in section \ref{sec:TB-MLCP}, and compare it with two existing methods, BR-MLCP and PS-MLCP, with some modifications for fair comparisons. We use one simulated dataset and two real datasets to compare the methods based on their length of prediction set and marginal coverage. We evaluate the results for a range of significance levels, including $\alpha$ = (0.02, 0.05, 0.08, 0.10, 0.12, 0.15, 0.20, 0.25, 0.30, 0.35). For each task, we implement the following 5 methods for comparisons:
 \begin{enumerate}
    \item \textit{Binary Relevance MLCP (BR-MLCP)}: With significance level = $\alpha/c$ for c labels.
     \item \textit{Power Set MLCP with present labelsets (PS-MLCP1)}: In this scenario, Power set MLCP is applied to the data with present labelsets and missing labelsets are not included in the prediction set.
    \item \textit{Power Set MLCP with all labelsets (PS-MLCP2)}: In this scenario, Power set MLCP is applied to the data with present labelsets %( i.e., 16 in our case), 
    and missing labelsets are added in the prediction set. %(i.e., $32 - 16=16$) 

    %( i.e., $16$ in our case) 
 
    \item \textit{Tree Based MLCP with present labelsets (TB1)}: The following two versions are implemented with present labelsets:
    \begin{enumerate}[label=(\roman*)]
\item  \textit{TB1 fixed-alpha} method: Using procedure 1 of FWER control i.e., $\alpha_i=\alpha/c$ with Algorithm \ref{alg:split_final} (omitting tuning set and $\lambda^*$).
\item \textit{TB1 adaptive-alpha}: Using procedure 2 of FWER control i.e., using optimal $\lambda^*$ with Algorithm \ref{alg:split_final}.
\end{enumerate}
    \item \textit{Tree Based MLCP with parent p-value labelsets (TB2)}: 
    The following two versions are implemented with parent $p$-value labelsets:
    \begin{enumerate}[label=(\roman*)]
    \item \textit{TB2 fixed-alpha}: Using procedure 1 of FWER control i.e., $\alpha_i=\alpha/c$ with Algorithm \ref{alg:split_final} (omitting tuning set and $\lambda^*$).
    \item \textit{TB2 adaptive-alpha}: Using procedure 2 of FWER control i.e., using optimal $\lambda^*$ with Algorithm \ref{alg:split_final}.
    \end{enumerate}

\end{enumerate}

For all experiments, we use the Naive-Bayes algorithm as the black-box algorithm, and the non-conformity score is computed as $1-$ predicted probability, as discussed in section \ref{cp}. We perform 50 replications of the simulations and real data and present the average results.
%One possible choice of weights is to use  $w_i = i/20$ for $i = 1, \ldots,  c-1,$ and $w_c = 1 - \sum_{i=1}^{c-1} w_i.$ This choice of weights was based on our empirical observations. It gives higher weights to the leaf layer than previous layers, which can be desirable in certain scenarios. We used the same weight settings for our simulation and real data experiments. 

\subsection{Simulation}

We generate n = 10,000 data points $(X_i, Y_i), i=1,2,\ldots,n,$ and $Y_i=(Y_i^1,\ldots, Y_i^5)$ denotes the response vector for $i^{th}$ observation with $c=5$ labels and each feature $X = (X_1, X_2)$ is derived  from Gaussian mixture model of two distributions as follows:

\begin{itemize}
    \item  $X_1 \sim 0.5* \mathcal{N}(1,0.3) + 0.5*\mathcal{N}(0,1)$
    \item  $X_2 \sim 0.3* \mathcal{N}(-1/2,0.2) + 0.7*\mathcal{N}(-\sqrt{3/2},0.4)$
\end{itemize}

\begin{figure}[h]
\centering
\includegraphics[width=0.6\textwidth]{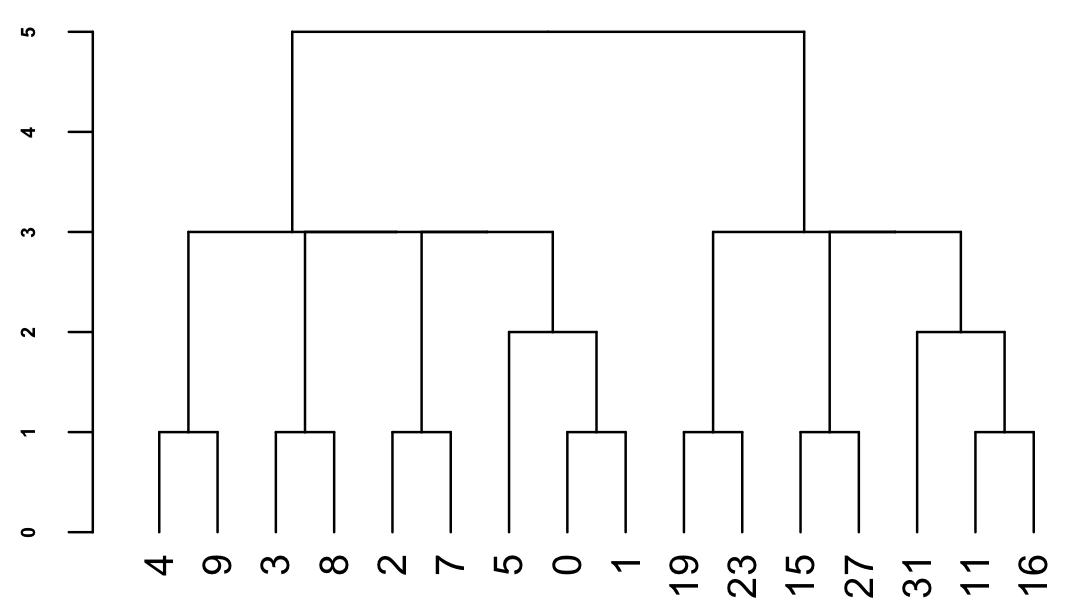}
\caption{Hierarchical tree for TB-MLCP (present labelsets) where $c=5$.}
\label{fig:label_present_sim}
\end{figure}

We  generate multiple labels (i.e., response variables) sequentially by $Y^1 \sim \text{Bernoulli}(\pi_1),$ where $\pi_1$ = $\mathbb{P}(Y^1|X) =  \dfrac{1}{1+e^{-z_1}}$ and $z_1 =  X^T\beta + \epsilon_1 $, for a coefficient vector $\beta= (\beta_0,\beta_1,\beta_2)=(2,2.5,2)$ and random error term, $\epsilon_1 = -0.5X_1^3.$ For consecutive labels, i.e., for the $j$-th label, $j>1$ and $j=2,\ldots,5$, we set $z_j = X^T\beta + \sum_{k=1}^{j-1} w_{jk} Y^k + \epsilon_j$, $w_{jk}=0$ for $k \geq j$, and $w_{jk}=1.5$ for $k<j$, indicating that the value of $(Y^1,\ldots, Y^{j-1})$ has a direct effect on $Y^j$. This is done to introduce dependence among the labels. The total number of possible labelsets is $2^c$, out of which 16 labelsets are present in the data as illustrated from the leaf nodes in Figure \ref{fig:label_present_sim}. However, \textit{TB2} (all labelsets) uses 32 possible labelsets to build the tree.

\begin{figure}
\centering
\includegraphics[width=0.7\textwidth]{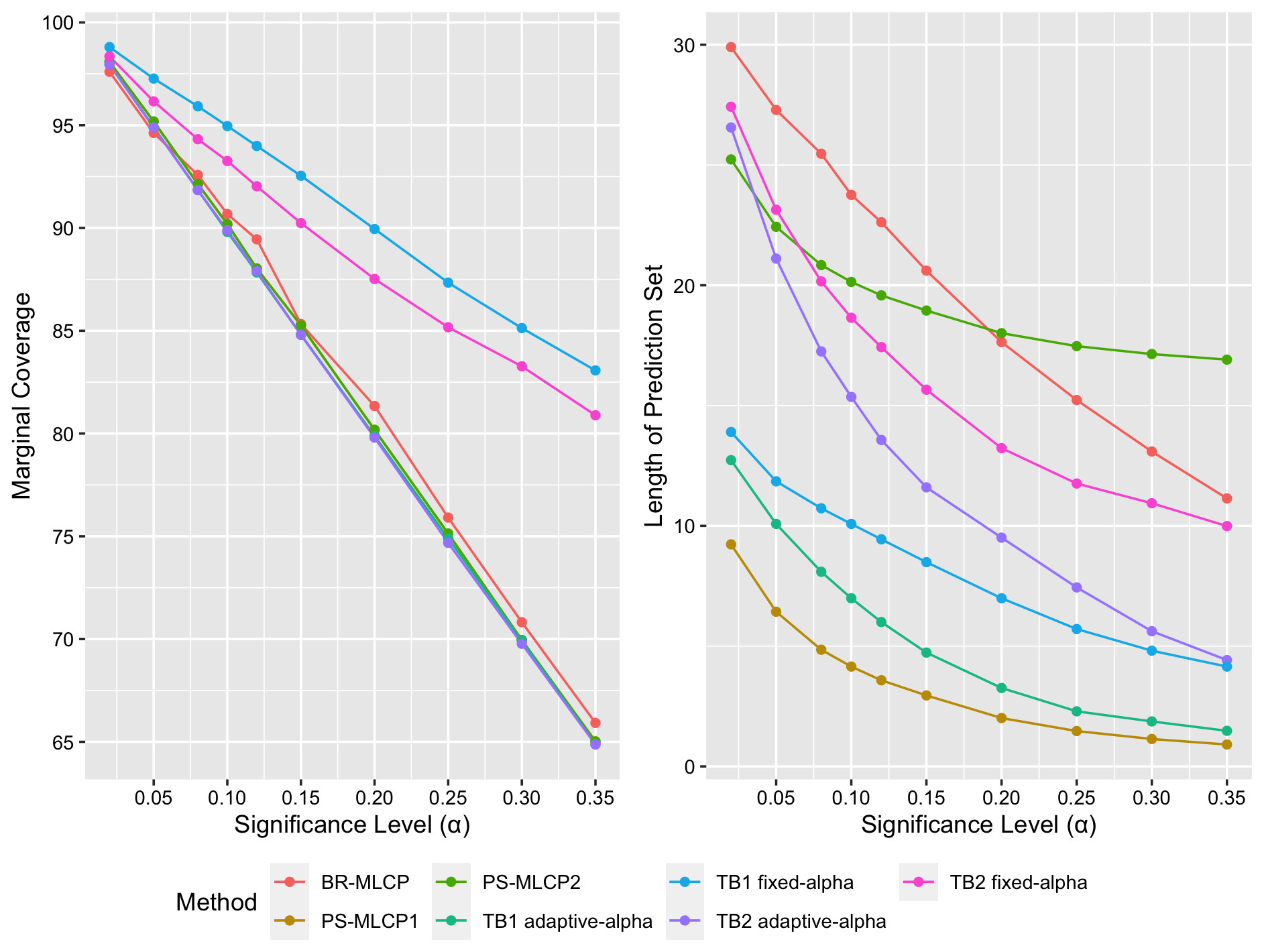}
\caption{Comparison results of our-proposed procedure with BR-MLCP and PS-MLCP in terms of length of the prediction set and marginal coverage.}
\label{fig:sim_res}
\end{figure}

In the adaptive-alpha approach, we split the data into four parts: proper training, calibration, tuning and test, with a split ratio of 30:30:20:20. In the fixed-alpha approach, we split the data into three parts: proper training, calibration and test, with a split ratio of 20:60:20.

In Figure \ref{fig:sim_res}, we present a comparison of our proposed procedures (\textit{TB1} and \textit{TB2} with both fixed-alpha and adaptive-alpha) with existing methods including BR-MLCP, PS-MLCP1, and PS-MLCP2. The left panel of Figure \ref{fig:sim_res} shows that all evaluated methods, including our proposed procedures, are able to achieve marginal coverage, indicating their ability to capture the true labels with high probability. \textit{TB1} and \textit{TB2} with fixed-alpha approach provide over-coverage, but with adaptive-alpha approaches, \textit{TB1} and \textit{TB2} provide nicely controlled coverages. The right panel of Figure \ref{fig:sim_res} indicates that BR-MLCP and PS-MLCP2 tend to generate larger prediction sets, while our proposed tree-based approach (\textit{TB2} with adaptive-alpha) provides shorter prediction sets than TB2 with fixed-alpha approach and PS-MLCP2 method, providing a more efficient solution to the problem of missing information than the PS-MLCP2 method. We also note that there is only a slight difference between the prediction set lengths of PS-MLCP1 and TB1 with adaptive-alpha. Also, the adaptive-alpha approach produces shorter prediction sets than the fixed-alpha approach in both \textit{TB1} and \textit{TB2} methods. 
%Overall, our proposed tree-based approach is expected to perform better than the PS-MLCP2 method in handling complex label dependencies and reducing the computational burden, owing to the nature of assigning parent $p$-value in the \textit{TB2} approach using the adaptive-alpha setting.

\subsection{Real Data Example}
The real datasets used in this study were obtained from the MULAN library.

\subsubsection{Scene Classification}\label{sec:scene}
The dataset pertains to the semantic classification of images into multiple labels, including beach, sunset, foliage, field, mountain, and urban, with a total of 6 labels. The dataset comprises 1,211 training and 1,196 test samples, each described by 294 features. Among the 64 possible labelsets in the data with $c=6$, only 14 were present, and we further filtered and preprocessed the data to focus on 8 of these labelsets (that occurred more than 20 times). To train and evaluate our proposed methods with existing methods,  we split training samples into proper training and calibration set as 40:60 split. If we require tuning set (i.e., for \textit{TB1} and \textit{TB2} with adaptive-alpha approach), then we split test samples into tuning and test with 50:50 split else, there is no split for test samples.

\begin{figure}[h]
\centering
\includegraphics[width=0.7\textwidth]{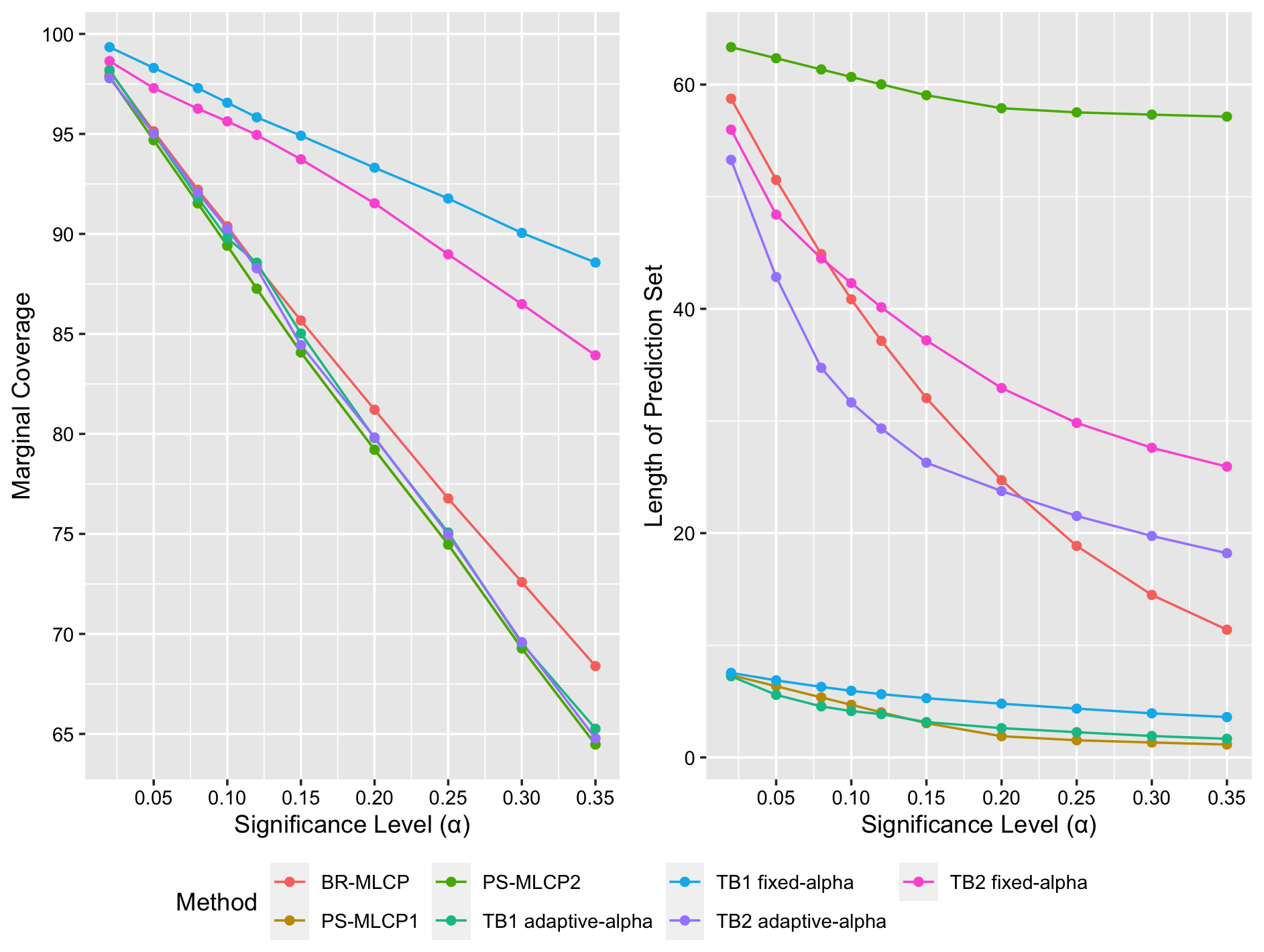}
\caption{Comparison results of our-proposed procedure with BR-MLCP and PS-MLCP in terms of length of the prediction set and marginal coverage on scene dataset.}
\label{fig:scene_res}
\end{figure}

Figure \ref{fig:scene_res} provides similar results as simulation studies. The left panel of the figure shows that all methods, including our proposed methods, provide marginal coverage guarantees with over coverage provided by the fixed-alpha approach. However, with adaptive-alpha approach, nice coverages are achieved for both \textit{TB1} and \textit{TB2} methods. The right panel of Figure \ref{fig:scene_res} shows that \textit{TB2} fixed-alpha and adaptive-alpha approach produces shorter sets compared to the PS-MLCP2 approach. Moreover, the set size produced by the \textit{TB2} adaptive-alpha approach is smaller than \textit{TB2} fixed-alpha approach. Meanwhile, PS-MLCP1 and \textit{TB1} adaptive-alpha produce similar set sizes. TB1 fixed-alpha set sizes are slightly larger than \textit{TB1} adaptive-alpha. BR-MLCP produces larger set sizes than other methods.

\subsubsection{Yeast}\label{sec:yeast}
 The dataset comprises 1,500 training and 917 test samples. Each gene is initially characterized by microarray expression data and phylogenetic profile, from which 103 features were extracted, and is linked to a subset of 14 functional classes (i.e., $c = 14$ labels).  Among the 16,384 possible labelsets in the data with $c=14$, only 16 were present. To train and evaluate our proposed methods, we have used the same setting as in section \ref{sec:scene}.

\begin{figure}[h]
\centering
\includegraphics[width=0.7\textwidth]{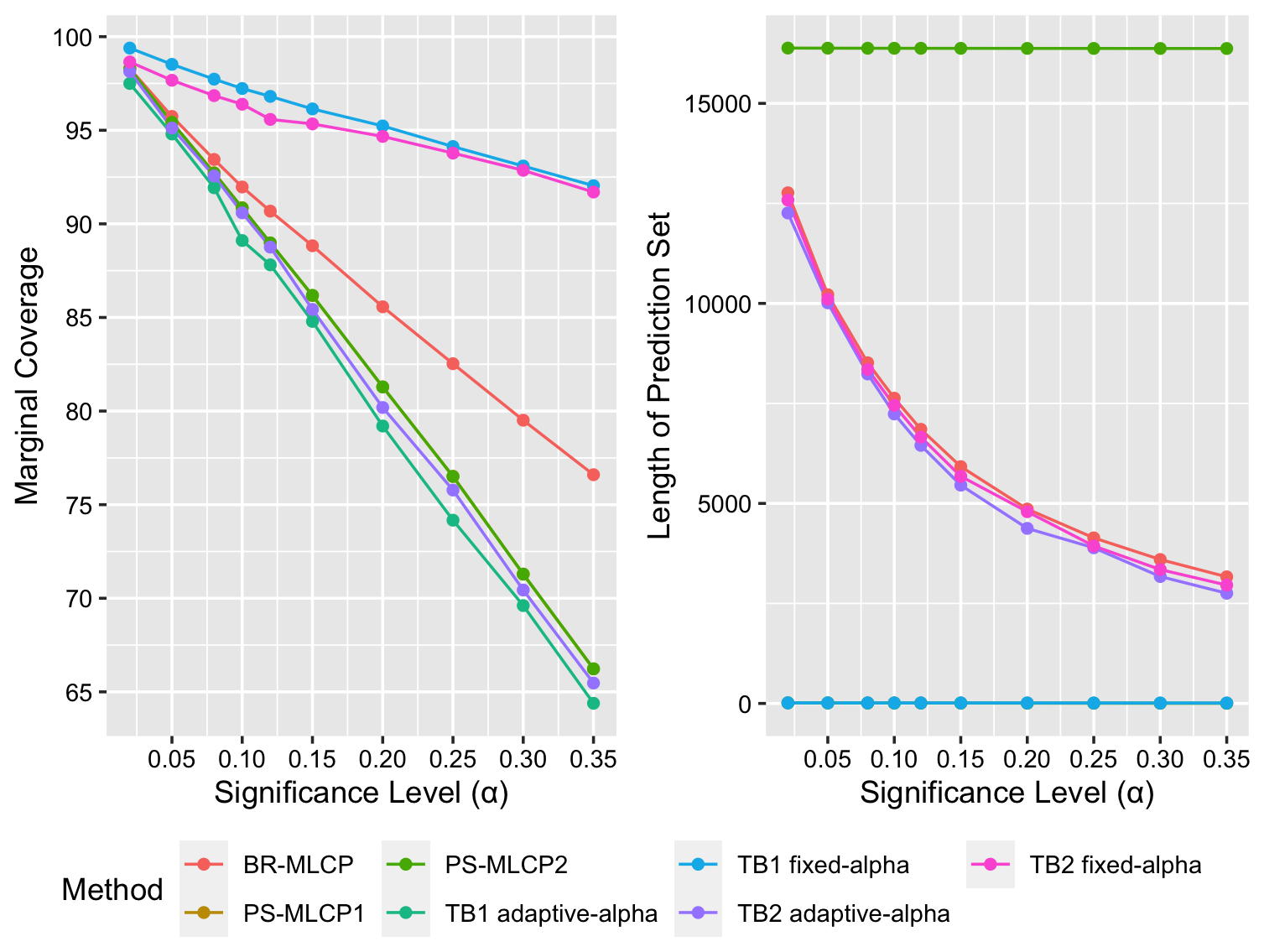}
\caption{Comparison results of our-proposed procedure with BR-MLCP and PS-MLCP in terms of length of the prediction set and marginal coverage on yeast dataset.}
\label{fig:yeast_result}
\end{figure}

The results presented in Figure \ref{fig:yeast_result} demonstrate that all of the methods under consideration offer marginal coverage guarantees. \textit{TB1} and \textit{TB2} with adaptive-alpha approach provides nice coverage as compared to fixed-alpha approach with both methods. In terms of prediction set size, our proposed method \textit{TB2} adaptive-alpha outperforms PS-MLCP2. Both PS-MLCP1 and \textit{TB1} adaptive-alpha show similar results in terms of the length of prediction sets. On the other hand, the length of the prediction set for BR-MLCP is shorter than that of PS-MLCP2. \textit{TB1} and \textit{TB2} with adaptive-alpha approach produce shorter set sizes as compared to \textit{TB1} and \textit{TB2} with fixed-alpha approach.

\subsubsection{Tuning Parameter}
\begin{figure}[h]
\centering
\includegraphics[width=0.65\textwidth]{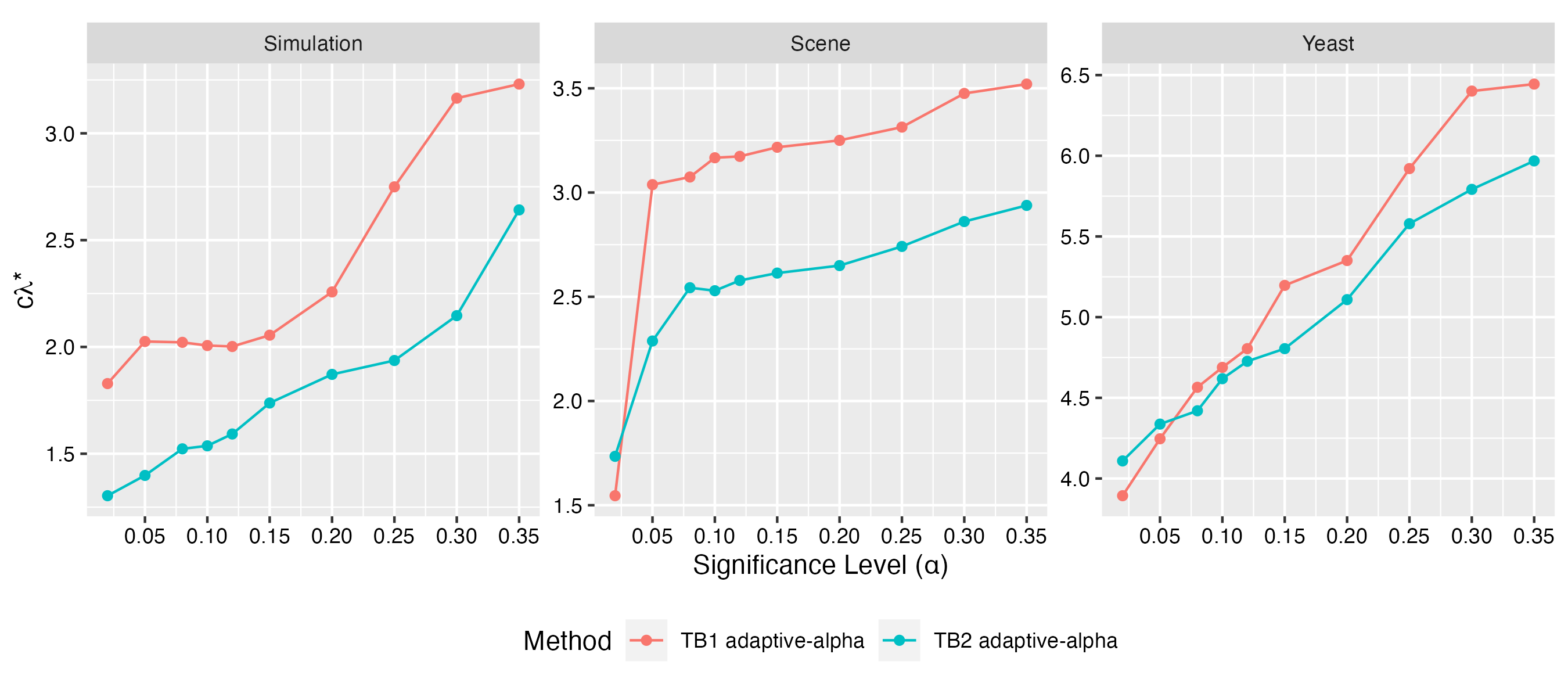}
\caption{Tuning parameter $\lambda^*$ behavior with \textit{TB1} and \textit{TB2} adaptive-alpha approaches in simulation, scene data and yeast data respectively.}
\label{fig:tune}
\end{figure}

In this section, we verify that the hyperparameter $\lambda^*$ is much larger than $1/c$, which implies that the adaptive-alpha approach is more powerful than the fixed-alpha approach in both \textit{TB1} and \textit{TB2} methods. This is because the adaptive-alpha approach uses higher critical value than the fixed-alpha approach. Figure \ref{fig:tune} illustrates $c \lambda^*$ as a function of $\alpha$ and demonstrates that $c \lambda^*$ is significantly greater than 1 for all three scenarios with $c=5,6,14$. This finding clearly explains why the adaptive-alpha approach performs better than the fixed-alpha approach in our simulation study and real data analysis.

\FloatBarrier

\section{Discussion}\label{sec:discussion}
In this paper, we proposed a tree-based method for multi-label classification problems using conformal prediction and multiple-testing. We presented four variants of the method, \textit{TB1} (present labelsets) and \textit{TB2} (all labelsets), both with fixed-alpha and adaptive-alpha approaches. Our method allows the use of any base classifier and provides prediction sets with a pre-specified coverage rate, such as $90\%$ while maintaining a small average length of the prediction set. We can tune the parameter $\lambda$ for adaptive-alpha approaches to ensure the nice marginal coverage guarantee. Our simulation study and real data analysis demonstrate that \textit{TB2} adaptive-alpha method outperforms other methods in terms of coverage and prediction set length and also accounts for missing label information. 
We anticipate that \textit{TB2} adaptive-alpha method will be a useful tool for researchers and practitioners working on multi-label classification problems. However, some critical questions still need to be addressed in future work. Firstly, we will explore more desired conditional coverage guarantees. Secondly, when a large number of test samples are considered, we may need to control some appropriate overall error rate, such as false discovery rate or false coverage rate. Additionally, it is necessary to extend the method to handle a larger number of labels and explore its performance in other types of machine-learning problems.

\bibliographystyle{plainnat}
\bibliography{main}

\newpage
\appendix

\section{Proofs}\label{app:proofs}

\subsection{Proof of Proposition 1: Validity of conformal $p$-values}

\begin{proof}
Suppose, for any given values of non-conformity scores, $v_1,\ldots,v_{n+1}$, they can be rearranged as:
    $$ \tilde{v}_1 < \ldots < \tilde{v}_k, $$
where each $\tilde{v}_i$ is repeated $n_i$ times such that $ \sum_{i=1}^k n_i=n+1.$ Let $E_v$ denote the event that the non-conformity scores ${\sigma_i}$, $i = 1, \ldots, n+1$ take on the specific values $v_1, \ldots, v_{n+1}$. Mathematically, 

$$E_v = \{ (\sigma_1, \sigma_2, \ldots, \sigma_{n+1}) : \sigma_i = v_i \text{ for some permutation of } v_1, \ldots, v_{n+1}\}$$

\noindent Since $\{X_i,Y_i\}_{i=1}^{n+1}$ are exchangeable, $\sigma_i$'s are also exchangeable, we have
\begin{equation}\label{eq:perm}
    \mathbb{P}\left\{ \sigma_{n+1} = \tilde{v}_j|E_v\right\} = \dfrac{n_j}{n+1}, \text{ for } j=1,\ldots,k
\end{equation}

\noindent Under the event of $E_v$ and $\sigma_{n+1} = \tilde{v}_j,$ we have from equation (\ref{eq:pvalue})
\begin{equation}\label{eq:p_march22}
    p(X_{n+1},Y_{n+1}) = \dfrac{\sum_{i=1}^{j-1} n_i + U_{n+1}\cdot n_j}{n+1}
\end{equation}

\noindent Thus, for any $t \in [0,1]$, we have from equation (\ref{eq:p_march22})
\begin{align*}
    \mathbb{P}\left\{ p(X_{n+1},Y_{n+1}) \leq t \big| E_v,\sigma_{n+1} = \tilde{v}_j)\right\} &= \mathbb{P}\left\{ \dfrac{\sum_{i=1}^{j-1}n_i + U_{n+1} \cdot n_j}{n+1} \leq t \big| E_v,\sigma_{n+1} = \tilde{v}_j\right\}\\
    &=
    \begin{cases}
    \dfrac{(n+1)t - \sum_{i=1}^{j-1}n_j}{n_j} & \text{ if } \dfrac{\sum_{i=1}^{j-1} n_i}{n+1} < t \leq \dfrac{\sum_{i=1}^{j} n_i}{n+1},\\
    0 & \text{ if } 0<t \leq \dfrac{\sum_{i=1}^{j-1} n_i}{n+1},\\
    1 & \text{ if } \dfrac{\sum_{i=1}^j}{n+1} < t \leq 1
    \end{cases}
\end{align*}
for $j=1,\ldots,k.$ Thus, for $\dfrac{\sum_{i=1}^{j-1} n_i}{n+1} < t \leq \dfrac{\sum_{i=1}^j n_i}{n+1} \text{ and } j=1,2,\ldots,k$ we have 

\begin{align*}
    \mathbb{P}\left\{ p(X_{n+1},Y_{n+1}) \leq t \mid E_v \right\} &=  
    \sum_{j=1}^k \mathbb{P} \left\{ p(X_{n+1},Y_{n+1}) \leq t \mid E_v,\sigma_{n+1} = \sigma_j\right\}\cdot \mathbb{P}\left\{\sigma_{n+1} = \sigma_j \mid E_v\right\}\\
    &= \sum_{j=1}^k   \dfrac{n_j}{n+1}\mathbb{P} \left\{ p(X_{n+1},Y_{n+1}) \leq t \mid E_v,\sigma_{n+1} = \sigma_j\right\}, \text{ using equation (\ref{eq:perm})}\\
    &=\dfrac{n_j}{n+1} \left[ \dfrac{(n+1)t - \sum_{i=1}^{j-1} n_i}{n_j}\right] + \dfrac{\sum_{i=1}^{j-1}n_i}{n+1}\\
    &= t - \dfrac{\sum_{i=1}^{j-1} n_i}{n+1} + \dfrac{\sum_{i=1}^{j-1}n_i}{n+1}\\
    &=t
\end{align*}
That is, 
\begin{equation}\label{eq:p_uni}
    p(X_{n+1},Y_{n+1}) \mid E_v \sim U(0,1)
\end{equation}
Taking the expectation on both sides of equation (\ref{eq:p_uni}), we have
$$p(X_{n+1},Y_{n+1}) \sim U(0,1)$$
This completes the proof.
\end{proof}

\subsection{Proof of Proposition 2}

\begin{proof} 
For the tree-based tested hypotheses formulated in section \ref{sec:mht}, there is only one true null hypothesis in each layer. Let $H_{i}$ denote the true null hypotheses for $i^{th}$ layer. Then, the FWER of the \textit{hierarchical testing procedure} with critical value $\alpha_i$ is given below.
\begin{align*}
        FWER &= \mathbb{P}\{ \text{at least one } H_{i} \text { is rejected for } i=1,\ldots,c \} 
          \end{align*}
Based on the construction of tested hypotheses, true null hypotheses $H_1,\ldots,H_c$ have the parent-children relationship (i.e., $H_{i}$ is a parent of $H_{i+1}$ for $i=1,\cdots,c-1$), we get
\begin{align*}
        FWER &=          
         \sum_{i=1}^c \mathbb{P}\{ H_{i} \text{ is rejected but } H_{j} \text{ are not rejected for } j=1, \ldots, i-1. \} \\
        &\leq  \sum_{i=1}^c \mathbb{P}\{p(X_{n+1},Y_{n+1}) \leq \alpha_i\} \text{ (based on the definition of hierarchical testing)}\\
        &= \sum_{i=1}^c \alpha_i, \text{ (by Proposition } \ref{prop:valid})\\
        & \leq \alpha,
  \end{align*}
the desired result. 

\end{proof}

\subsection{Proof of Proposition 5}

\begin{proof} 
Let $\mathcal{D}_{tune}$ be the tuning dataset with $\lvert \mathcal{D}_{tune} \rvert = n$. By Algorithm \ref{alg:algo}, we have
\begin{equation}\label{eq:*}
    \dfrac{1}{n}\sum_{j=1}^n \mathbb{I} \{Y_j \in \hat{C}_{tune} (X_j,\alpha^*)\} = 1- (1+\dfrac{1}{n})(\alpha-\dfrac{1}{n})
\end{equation} 
Note that for any observation $(X_j,Y_j) \in \mathcal{D}_{tune},$ the event of $Y_j \in \hat{C}_{tune} (X_j,\alpha^*)$ is equivalent to  the corresponding true null hypothesis is not rejected for each layer $i$, which in turn is equivalent to $p^{i}({X}_j,Y_j) \geq \alpha^*$, where $p^{i}({X}_j,Y_j)$ denotes the smoothed conformal $p$-value for $i^{th}$ layer and $j^{th}$ observation as defined in equation (\ref{eq:pvalue}). Define $p(X_j,Y_j) = \underset{{i \in \{1,\ldots,c\}}}{\min}\{p^{i}({X}_j,Y_j)\}$. Thus, 
$\mathbb{I} \{Y_j \in \hat{C}_{tune}(X_j,\alpha^*\} = \mathbb{I}\{p(X_j,Y_j) \geq \alpha^*\}$. Therefore, equation (\ref{eq:*}) is equivalent to
\begin{equation}
    \dfrac{1}{n} \sum_{j=1}^n \mathbb{I}\{p(X_j,Y_j) \geq \alpha^*\} = 1- (1+\dfrac{1}{n})(\alpha-\dfrac{1}{n})
\end{equation}
 Thus, $\alpha^* = \hat{Q}_n((1+\dfrac{1}{n})(\alpha-\dfrac{1}{n}))$, the upper $(1-(1+\dfrac{1}{n})(\alpha-\dfrac{1}{n}))$ quantile of $\{p(X_j,Y_j)\}_{j=1}^n.$

 \noindent Let $Z_j = p(X_j,Y_j),$ $j=1,\ldots,n$ and $Z_{n+1} = \underset{{i \in \{1,\ldots,c\}}}{\min}\{p^{i}(X_{n+1},Y_{n+1})\}$. Note that $(X_i,Y_i)$'s are exchangeable, so $Z_1,\ldots,Z_{n+1}$ are also exchangeable. By Lemma \ref{lemma:label1}, we have
 $$\mathbb{P}\{Z_{n+1} \leq \hat{Q}_n((1+\dfrac{1}{n})(\alpha-\dfrac{1}{n}))\} \geq \alpha-\dfrac{1}{n}$$

\noindent By using the fact of $\alpha^* = \hat{Q}_n((1+\dfrac{1}{n})(\alpha-\dfrac{1}{n}))$, we have

\begin{equation}\label{eq:label**}
\mathbb{P}\{Z_{n+1} \leq \alpha^*\} \geq \alpha-\dfrac{1}{n}
\end{equation}
By equation (\ref{eq:pvalue}),  $Z_1,\ldots,Z_{n+1}$ are almost surely distinct, then Lemma \ref{lemma:label1} also gives
\begin{equation}\label{eq:label***}
    \mathbb{P}\{Z_{n+1} \leq \alpha^*\} \leq \alpha 
\end{equation}

 \noindent Note that FWER of procedure 2 is given by $\mathbb{P}\{Z_{n+1} \leq \alpha^*\}$. By equation (\ref{eq:label**}) and (\ref{eq:label***}), we have
 \iffalse
\begin{equation}
    \alpha \leq FWER  \leq \alpha+ \dfrac{1}{n}
\end{equation}
  Let $\alpha= \alpha-\dfrac{1}{n},$ we have
 \fi
  \begin{equation}\label{eq:pr2l_b}
      \alpha-\dfrac{1}{n} \leq FWER \leq \alpha
 \end{equation} 
This completes the proof.
\end{proof}

\end{document}